# Temporal scaling of the growth dependent radiative properties of microalgae

J. M. Zhao[a],*, C. Y. Ma[a], L. H. Liu[a],*

[a] *School of Energy Science and Engineering, Harbin Institute of Technology, 92 West Dazhi Street, Harbin*

*150001, People's Republic of China*

## Abstract

The radiative properties of microalgae are basic parameters for analyzing light field distribution in photobioreactors (PBRs). With the growth of microalgae cell, their radiative properties will vary with growth time due to accumulation of pigment and lipid, cell division and metabolism. In this work, we report both experimental and theoretical evidence of temporal scaling behavior of the growth dependent radiative properties of microalgae cell suspensions. A new concept, the temporal scaling function (TSF), defined as the ratio of absorption or scattering cross-sections at growth phase to that at stationary phase, is introduced to characterize the temporal scaling. The temporal evolution and temporal scaling characteristics of the absoroption and scattering cross-sections of three example microalgae species, *Chlorella vulgaris*, *Chlorella pyrenoidosa*, and *Chlorella protothecoides,* were experimentally studied at spectral range 380 to 850 nm. It is shown that the TSFs of the absorption and scattering cross-sections for different microalgae species are approximately constant at different wavelength, which confirms theoretical predictions very well. With the aid of the temporal scaling relation, the radiative properties of microalgae at any growth time can be calculated based on those measured at stationary phase. The findings of this work will help the understanding of time dependent radiative properties of microalgae and facilitate their applications in light field analysis in PBRs design.

___________________________________________________________________

*Corresponding author. Tel.: +86-451-86412138.

E-mail addresses*: jmzhao@hit.edu.cn (J.M. Zhao), lhliu@hit.edu.cn (L. H. Liu).

**Highlights**

- Finding of temporal scaling behavior of the radiative properties of microalgae

- Both theoretical proof and experimental examination are presented

- Concept of temporal scaling function(TSF) is introduced to characterize temporal scaling

- Temporal scaling relation shows TSF is wavelength independent

- Radiative properties at any time can be obtained using those at stationary phase based on TSF

## 1. Introduction

The world is facing an unprecedented combination of economic and environmental challenges to meet the huge energy demand [1]. The extensive use of fossil fuels leads to growing greenhouse gases emissions which then influence the global climate [2]. The primary energy consumption of the world was predicted to increase by 37% between 2013 and 2035 [3]. Hence, it is crucial to find renewable and clean fuels to replace the traditional fossil fuels. Microalgae is a kind of fast growing microorganisms on the earth, which can produce oil of per unit area 7-31 times greater than terrestrial plants [4], and hence is considered to be the most promising alternative resources for biofuel production. Ullah et al. [4] reported that biofuels have the potential to meet 50% of the world energy consumption, meanwhile without net emissions of carbon dioxide. Moreover, the cultivation of microalgae can produce other value-added by-products which make the process more economical [5-7].

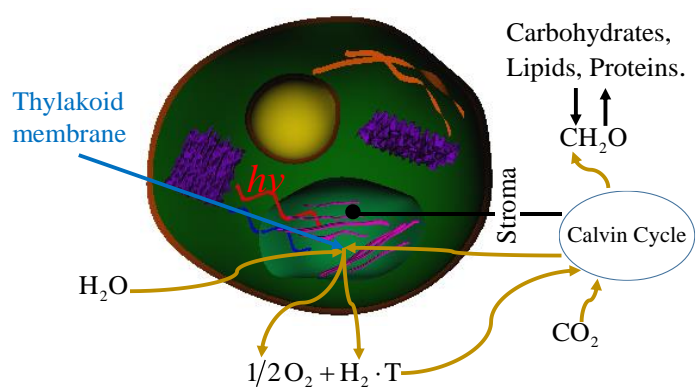

**FIG. 1.** Schematic of photosynthetic process in chloroplasts,
light absorption and water splitting is on thylakoid membrane *via* photochemical electron transport
and the stroma for dark reactions. "T" represents the chemically binds hydrogen [8].

Microalgae can produce carbohydrates, proteins, lipids and oxygen within the cells, some species can also produce $H_2$ through the photosynthesis process as illustrated in Fig. 1. The photosynthesis process consists of two reactions, namely, light and dark reactions. During light reactions, photons are absorbed in chloroplasts to produce ATP and NADPH. These products are used in the dark reactions to produce carbohydrates, proteins and lipids [8]. Microalgae typically have efficiencies about 10 times of the terrestrial higher plants at converting light energy into bioenergy per unit area [9, 10], due to microalgae cells growing in aqueous media. Many species of microalgae can also utilize waste water for microalgae cultivation [11]. Moreover, the microalgae can be cultivated in open or closed PBRs without occupying the agriculture production lands [7, 11].

Although the cultivation of microalgae for biofuel production presents many advantages, its cultivation cost is still high in PBRs. It is still facing numerous challenges, such as, light utilization efficiency, a large amount of nutrients and auxiliary energy requirement during the cultivation of microalgae [12-14]. Light utilization efficiency of the PBRs significantly affects the microalgae biomass productivity [15, 16]. Indeed, photosynthesis process of microalgae requires an optimum irradiance to achieve their maximum biomass production. The microalgae will suffer light inhibition for excessive irradiance [17], meanwhile, limited light irradiance will cause the decrease of biomass productivity [18, 19]. Hence careful light transfer analysis must be conducted to design and optimize the light field distribution in PBRs to make them more efficient.

The radiative properties of microalgae are basic parameters for obtaining light field in PBRs. Radiative properties of many microalgae species at the VIS-NIR spectral range has already been experimentally determined [20-24]. Until recently, most reported radiative properties of microalgae were measured at the stationary growth phase, partly due to the difficulty in on-line measurement. However, with the growth of microalgae cell, their radiative properties will vary with time due to accumulation of pigment and lipid, cell division and metabolism. The effect of cell growth on the radiative properties of microalgae has been rarely studied. Most recently, Heng and Pilon [25] reported their experimental study of the temporal evolution radiation characteristics of absorption and scattering cross-sections of *Nannochloropsis oculata* in full growth phase during batch cultivation. The *Nannochloropsis oculata* was grown in a flat-plate PBR under constant irradiance by red LEDs emitting at 630nm. The radiation characteristics between 400 and 750 nm and pigment concentrations were measured every 24 hours for up to 18 days. They found that the absorption and scattering cross-sections of the microalgae varied significantly with growth time in response to change in light fields and nutrients availability.

In this study, we report both experimental and theoretical evidence of the temporal scaling behavior of the growth dependent radiative properties of microalgae cell suspensions. The concept of temporal scaling function defined as the ratio of absorption or scattering cross-sections at growth phase to that at stationary phase is introduced to characterize the temporal scaling, which is proved to be wavelength independent. The temporal scaling characteristics of the scattering and absorption cross-sections of three example microalgae, *Chlorella vulgaris*, *Chlorella pyrenoidosa*, and *Chlorella protothecoides*, were experimentally studied at spectral range 380 to 850 nm.

## 2. The radiative properties of microalgae

The microalgae suspensions in PBRs is a typical kind of participating media of radiative transfer. When a light beam transports in the microalgae suspensions, it encounters scattering and absorption by the microalgae cells as well as by the bubbles and culture medium. The governing equation for light transport within the microalgae suspensions is the radiative transfer equation, which can be written as [26]

$$\mathbf{s} \cdot \nabla I_\lambda(\mathbf{r}, \mathbf{s}) + \beta_\lambda I_\lambda(\mathbf{r}, \mathbf{s}) = \frac{\kappa_{s,\lambda}}{4\pi} \int_{4\pi} I_\lambda(\mathbf{r}, \mathbf{s}') \Phi_\lambda(\mathbf{s}', \mathbf{s}) \mathrm{d}\Omega' \tag{1}$$

where $I_\lambda$ is the spectral radiative intensity ($\mathrm{W/m^2 \cdot nm \cdot sr}$) in direction $\mathbf{s}$ and at location $\mathbf{r}$, $\kappa_{s,\lambda}$ is the scattering coefficient ($\mathrm{m^{-1}}$), $\beta_\lambda = \kappa_{a,\lambda} + \kappa_{s,\lambda}$ is the extinction coefficient ($\mathrm{m^{-1}}$), $\kappa_{a,\lambda}$ is the absorption coefficient ($\mathrm{m^{-1}}$), $\Phi_\lambda(\mathbf{s}', \mathbf{s})$ is the scattering phase function and $\Omega'$ is the solid angle. The scattering phase function $\Phi_\lambda(\mathbf{s}', \mathbf{s})$ stands for the probability that the radiation transfer in the solid angle $\mathrm{d}\Omega'$ around the direction $\mathbf{s}'$ scattered into the solid angle $\mathrm{d}\Omega$ around the direction $\mathbf{s}$ and is normalized with the following equation [26]

$$\frac{1}{4\pi}\int_{4\pi}\Phi_\lambda(\mathbf{s}',\mathbf{s})\mathrm{d}\Omega = 1 \tag{2}$$

The basic characteristics of the scattering phase function is given by the asymmetry factor $g_\lambda$, which describes isotropic, backward or forward scattering features, is defined as [26]

$$g_\lambda = \frac{1}{4\pi}\int_{4\pi}\Phi_\lambda(\mathbf{s}',\mathbf{s})\cos\theta\,\mathrm{d}\Omega \tag{3}$$

where $\theta$ is the angle between the incident direction $\mathbf{s}'$ and scattering direction $\mathbf{s}$. Previous studies showed that the microalgae cells featured strongly forward scattering with $g_\lambda$ values around 0.97 [27]. Generally, the absorption and scattering coefficients of microalgae suspensions are time-dependent due to the growth of microalgae cells. When neglecting the contributions of bubbles and culture medium, they can be expressed in terms of the average absorption cross-section $C_{\mathrm{abs},\lambda}$ and average scattering cross-section $C_{\mathrm{sca},\lambda}$ ( m$^2$/#) as [26]

$$\kappa_{a,\lambda}(t) = C_{\mathrm{abs},\lambda}(t)N(t) \tag{4}$$

$$\kappa_{s,\lambda}(t) = C_{\mathrm{sca},\lambda}(t)N(t) \tag{5}$$

respectively, where $N(t)$ is the cell number density (#/m$^3$) as a function of growth time $t$. The number density is difficult to be measured for multicellular microalgae. Alternatively, the radiative properties can also be expressed in terms of the average mass absorption cross-section $C_{\mathrm{abs},\lambda}^{\mathrm{M}}$, average mass scattering cross-section $C_{\mathrm{sca},\lambda}^{\mathrm{M}}$ (m$^2$/kg) and mass concentration $X$ (kg/m$^3$) [27] as

$$\kappa_{a,\lambda}(t) = C_{\mathrm{abs},\lambda}^{\mathrm{M}}(t)X(t) \tag{6}$$

$$\kappa_{s,\lambda}(t) = C_{\mathrm{sca},\lambda}^{\mathrm{M}}(t)X(t) \tag{7}$$

During microalgae cell growing, it will experience the accumulation of pigment and lipid, cell division and metabolism. The time dependent cell concentration of microalgae can be modeled based on the general exponential growth law as [1, 28]

$$\frac{\mathrm{d}X(t)}{\mathrm{d}t} = \mu(t)X(t) \tag{8}$$

where the $\mu$ is the specific growth rate (h$^{-1}$) characterizing the growth kinetics, which is generally a function of growth time. Analytical solution of Eq. (8) can be obtained as

$$X = X_0 \exp\left[\tau_X(t)\right] = X_0 \exp\left[\int_0^t \mu(t)\,\mathrm{d}t\right] \qquad (9)$$

where $X_0$ stands for the initial cell concentration (at $t=0$) and $\tau_X(t) = \int_0^t \mu(t)\,\mathrm{d}t$.

## 3. Theoretical proof of the temporal scaling relation of radiative properties

Previous study showed that the radiative properties of microalgae varied significantly with growth time [29]. However, there is still no theoretical models proposed to describe the time-dependent radiative properties of microalgae. Since the state of a cell relies on previous state in a growing process, there must be some connections between the radiative properties of microalgae at the different growth period. The connections are still to be revealed. Here, as an initial attempt in this direction, a temporal scaling relation of the radiative properties of microalgae at different growth period is established.

For the ease of following derivation, the wavelength dependence ($\lambda$) of radiative properties is expressed explicitly in the function argument. Firstly, we consider the temporal scaling relation of absorption cross-section of microalgae. It is well known that the spectral absorption coefficient of microalgae can be expressed as [30-32]

$$\kappa_a(\lambda) = \sum_i E_{a,i}(\lambda) X_{\mathrm{pig},i} \qquad (10)$$

where $E_{a,i}(\lambda)$ (m$^2$/kg) is the specific spectral absorption cross-section of the $i$-th pigment, and $X_{\mathrm{pig},i}$ is the mass concentrations (kg/m$^3$) of the $i$-th pigment. The main pigments in microalgae cells include chlorophyll a, b and c, and $\beta$-carotenoid, etc., and their specific absorption cross-sections were reported by Bidigare et al. [31]. Since the radiative properties of pigments are constant, the time-dependent spectral absorption coefficient can be written as

$$\kappa_a(\lambda,t) = \sum_i E_{a,i}(\lambda) X_{\mathrm{pig},i}(t) \qquad (11)$$

which indicates the variation of cell absorption coefficient with growth time is influenced by the accumulation of pigments in each cell. Supposing the average pigment concentration of a single cell for the $i$-th pigment is $x_{\text{pig},i}$ (kg/#), then

$$X_{\text{pig},i}(t) = x_{\text{pig},i}(t)\,N(t) \tag{12}$$

Combine Eq. (11) and Eq. (12), the average absorption cross-section of microalgae can be written as

$$C_{\text{abs}}(\lambda, t) = \kappa_a(\lambda, t)\,/\,N(t) = \sum_i E_{a,i}(\lambda) x_{\text{pig},i}(t) \tag{13}$$

Here, it is assumed that the accumulation of pigments in each cell with growth time follows the general exponential growth law as

$$\frac{\mathrm{d}x_{\text{pig},i}}{\mathrm{d}t} = \mu_p x_{\text{pig},i} \tag{14}$$

where $\mu_p$ denotes the average accumulation rate of pigments in each cell, which generally will be a function of time. The solution of Eq. (14) yields

$$x_{\text{pig},i}(t) = x_{\text{pig},i}(0) e^{\tau_p(t)} \tag{15}$$

where $\tau_p(t) = \int_0^t \mu_p(t)\,\mathrm{d}t$. For different pigments, the average accumulation rates are assumed to be the same, while their initial concentrations in each cell may be different.

Substitution of Eq. (15) into Eq. (13) leads to

$$C_{\text{abs}}(\lambda, t) = e^{\tau_p(t)} \sum_i E_{a,i}(\lambda) x_{\text{pig},i}(0) \tag{16}$$

Taking the time at stationary phase $t_{\text{STP}}$ as a reference time, the absorption cross-section is

$$C_{\text{abs}}(\lambda, t_{\text{STP}}) = e^{\tau_p(t_{\text{STP}})} \sum_i E_{a,i}(\lambda) x_{\text{pig},i}(0) \tag{17}$$

Dividing Eq. (16) by Eq. (17) yields

$$\frac{C_{\text{abs}}(\lambda, t)}{C_{\text{abs}}(\lambda, t_{\text{STP}})} = e^{\tau_p(t) - \tau_p(t_{\text{STP}})} \tag{18}$$

This is the main result of this section and also the main finding of present study, namely, *the ratio of time-dependent absorption cross-section of microalgae at any time to that at stationary phase is wavelength independent,* which is a temporal scaling relation for radiative properties of microalgae. As will be shown later, the predicted temporal scaling relation is confirmed by experiments very well.

As a potential application, the temporal scaling relation will significantly simplify the modeling of time-dependent radiative properties of microalgae. By introducing a *temporal scaling function* as $Z_a(t, t_{\text{STP}}) = e^{\tau_p(t) - \tau_p(t_{\text{STP}})}$, if it has already been determined (such as by experiment), the radiative properties at any growth time can be just calculated based on the radiative properties measured at the stationary phase from

$$C_{\text{abs}}(\lambda, t) = C_{\text{abs}}(\lambda, t_{\text{STP}}) \, Z_a(t, t_{\text{STP}}) \tag{19}$$

Note that the time-dependence function $Z_a(t, t_{\text{STP}})$ may be different for different microalgae species, which can be considered as a basic growth property for the microalgae.

The temporal scaling relation has already been established for the absorption cross-sections $C_{\text{abs}}(\lambda, t)$, how about the scattering cross-sections $C_{\text{sca}}(\lambda, t)$? By analogy with the absorption cross-sections of microalgae cells, supposing that there are some scatterers (cell walls, internal structures and inhomogeneity of each cell) that determines the spectral scattering characteristics of the microalgae cells, their concentration also varies with cell growth, similar to the pigments contribution to light absorption, such that the scattering coefficient can be written similar to Eq. (11) as

$$\kappa_s(\lambda, t) = \sum_i E_{s,i}(\lambda) X_{\text{scat},i}(t) \tag{20}$$

here $E_{s,i}(\lambda)$ (m$^2$/kg) and $X_{\text{scat},i}$ (kg/m$^3$) are introduced as the specific spectral scattering cross-section and the mass concentrations of the *i*-th scatterer, respectively. Following the similar derivation procedure presented above, a *temporal scaling function* for scattering cross-section $Z_s(t, t_{\text{STP}})$ can be defined and the temporal scaling for scattering cross-section can be written as

$$C_{\text{sca}}(\lambda, t) = C_{\text{sca}}(\lambda, t_{\text{STP}}) \, Z_s(t, t_{\text{STP}}) \qquad (21)$$

Though the derivation of the temporal scaling relation for scattering cross-section is less rigid than that of the absorption cross-section, further experimental results still show good temporal scaling relation of the scattering cross-section of microalgae, which indicates reasonability of the assumptions.

In summary of this section, temporal scaling relation of the absorption cross-sections and scattering the cross-sections of microalgae at different growth period are established. By using the temporal scaling relation, the radiative properties at any growth time can be calculated based on the radiative properties measured at the stationary phase.

## 4. Experimental methods

Experimental examination of the temporal scaling of the radiative properties of microalgae was conducted. The spectral absorption cross-sections and scattering cross-sections of three kinds of microalgae were measured at different growth time. In this section, the cultivation procedure of the example microalgae and the experimental method for measuring the cell number density, spectral extinction and absorption coefficients of microalgae suspensions are presented.

### 4.1 Cultivation and sample preparation

The microalgae species of *Chlorella vulgaris*, *Chlorella pyrenoidosa*, *Chlorella protothecoides* were selected as examples, which were purchased from the Freshwater Algae Culture Collection at the Institute of Hydrobiology (FACHB) located in Wuhan, China. They were cultivated in BG11 medium in 250ml culture bottles fitted with vented caps and exposed to continuous luminous flux of 4500-5000 Lux provided by fluorescent light bulbs, respectively. The composition of the BG11 medium are (per liter of the distilled water): $NaNO_3$ 1.5 g, $K_2HPO_4$ 0.04 g, $MgSO_4 \cdot 7H_2O$ 0.075 g, $CaCl_2 \cdot 2H_2O$ 0.036 g, Citric acid 0.006 g, Ferric ammonium citrate 0.006 g, $EDTANa_2$ 0.001 g, $Na_2CO_3$ 0.02 g and Trace metal solution A5 (1 ml). One liter of the trace metal solution A5 contains 2.86 g $H_3BO_3$, 1.86 g $MnCl_2 \cdot 4H_2O$, 0.22 g $ZnSO_4 \cdot 7H_2O$, 0.39 g

Na$_2$MoO$_4$·2H$_2$O, 0.08 g CuSO$_4$·5H$_2$O, 0.05 g Co(NO$_3$)$_2$·6H$_2$O. The PH of BG11 medium was adjusted to 7.5.

Fig. 2 shows the optical micrograph of *Chlorella vulgaris*, *Chlorella pyrenoidosa*, and *Chlorella protothecoides*, respectively. As shown, the three species of microalgae are unicellular and approximately spherical shaped.

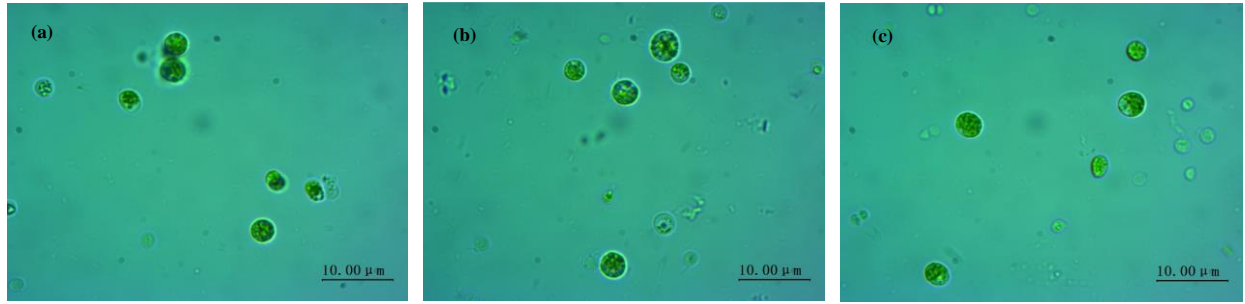

**FIG. 2.** Optical micrographs of three microalgae species,
(**a**) *Chlorella vulgaris*, (**b**) *Chlorella pyrenoidosa*, and (**c**) *Chlorella protothecoides*, respectively.

After 12 days of cultivation under continuous fluorescent light, 20 ml of the culture diluted with 320 ml of BG11 medium, then each 170 ml of the diluted culture was transferred into a 2 cm thick flat-plate PBR suitable for measuring the spectral extinction coefficient and absorption coefficient with the method described in Section 4.2. The PBR was placed into the incubator at the cultivation condition of dark and light cycle 12 and 12 h, and exposed to an illuminance of 4500-5000 Lux from fluorescent light bulbs at the light cycle. The temperature in the incubator was kept constant at 25 ℃.

The cell number density of the culture was counted using a Petroff-Hausser counting chamber (Hausser scientific, USA). The cell size distribution was measured based on micrographs taken from a biological microscope (UB203i-5.0M, China) connected to a CCD camera. The ImageJ software (developed at the National Institutes of Health, http://rsb.info.nih.gov/ij) was used to manually measure the major and minor diameters of individual cell. All the measurements were carried out at room temperature (about 25 ℃). The cell number density was measured from two samples and each was counted twice.

*4.2 Measurement methods of the radiation characteristics*

The spectral extinction coefficient of the microalgae suspensions was obtained from the normal-normal transmittance ($T_{n,\lambda}$) measurements. The measurements were performed in a 2 cm pathlength cuvettes using a V–VASE ellipsometer (J.A.Woollam Company, USA) at the spectral range 380 to 850 nm as illustrated in Fig. 3(a). Before the measurement, the microalgae suspensions was well mixed to ensure uniform distribution of the cells. The distance between detector and cuvette was 200 mm and the diameter of the detector was 2 mm, which was demonstrated to give enough accurate result for the extinction coefficient in our previous study [24, 33]. Based on the measured normal-normal transmittance, the spectral extinction coefficient $\beta_\lambda$ was calculated from [1]

$$\beta_\lambda = -\frac{1}{L}\ln\frac{T_{n,\lambda}}{T_{n,\lambda,ref}} \qquad (22)$$

where $T_{n,\lambda,ref}$ is the measured transmittance using the reference medium (here the BG11 medium).

The spectral absorption coefficient $\kappa_{a,\lambda}$ was obtained from normal-hemispherical transmittance ($T_{h,\lambda}$) measurements, performed with an integrating sphere at the spectral range 380 to 850 nm as illustrated in Fig. 3(b), and calculated from [1]

$$\kappa_{a,\lambda} = -\frac{1}{L}\ln\frac{T_{h,\lambda}}{T_{h,\lambda,ref}} \qquad (23)$$

where $T_{h,\lambda,ref}$ is the measured normal-hemispherical of the reference medium (here the BG11 medium). For measuring the absorption coefficient of microalgae, it is common to omit the back scattering of microalgae due to the strong forward scattering of microalgae cells (with asymmetry factor about 0.97 [27, 34]) that most of the incident photons are scattered into the forward direction.

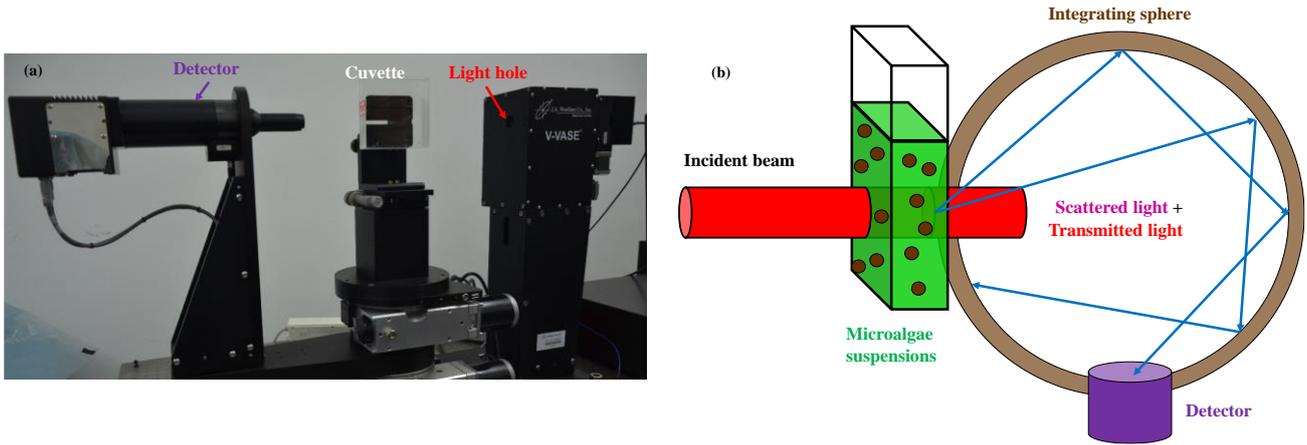

**FIG. 3.** Experimental techniques to measure the spectral extinction coefficient and the absorption coefficient of microalgae. **(a)** experimental setup for measuring the extinction coefficient, **(b)** schematic for measuring the absorption coefficient using integrating sphere.

Each measurement was performed for two samples with reference to each other. The normal-normal and normal-hemispherical transmittances measurement were performed three times and the averaged results are presented. The average absorption and extinction cross-sections $C_{\text{abs},\lambda}$ and $C_{\text{ext},\lambda}$ were obtained based on the absorption and extinction coefficients $\kappa_{a,\lambda}$ and $\beta_\lambda$ according to Eq. (4). The average scattering cross-section $C_{\text{sca},\lambda}$ was then calculated from $C_{\text{sca},\lambda} = C_{\text{ext},\lambda} - C_{\text{abs},\lambda}$.

## 5. Results and discussion

The measured growth curves of cell number density and cell size distribution data are presented for the three microalgae species. The average absorption and scattering cross-sections of different microalgae samples were measured every 24 hours up to 18 days. The time dependent spectral absorption and scattering cross-sections of the three microalgae species were measured in the spectral range 380 to 850 nm. The temporal scaling of the spectral absorption and scattering cross-sections of microalgae were analyzed. The experimental determined TSFs for the three microalgae species are presented.

### 5.1 Growth characteristics and time dependent radiative properties

The growth curve of cell number density shows basic growth characteristics of microalgae. Figure 4 shows the temporal evolution of the cell number density $N(t)$ for the three species of microalgae cultures.

Each data point in the figure gives the arithmetic mean value of multiple measurements. As shown in the figure, the curves show typical growth phases for microalgae, i.e. (1) the lag phase, (2) the exponential phase, and (3) the stationary phase. The lag phase was characterized by slow initial growth rates and occurred between 0 and 20 hours. It is attributed to the microalgae's adaptation to the new growth conditions after they were transferred from the culture bottle to the flat-plate PBR. The exponential growth phase followed the lag phase and was characterized by large growth rates. The stationary phase occurred after approximately 287 hours.

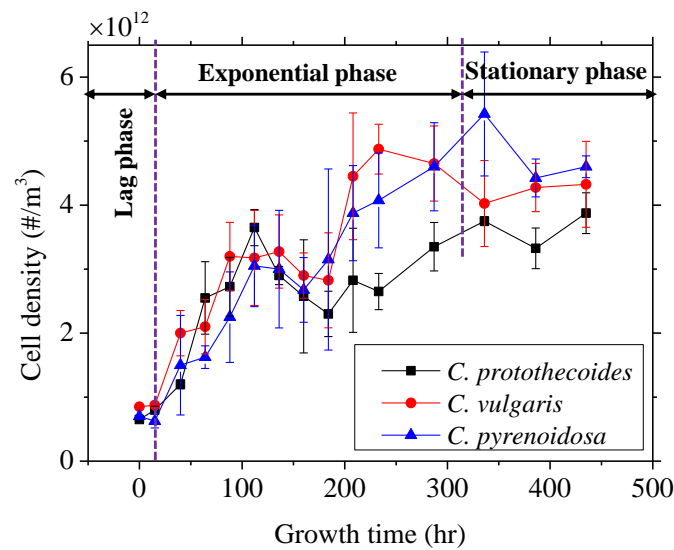

**FIG. 4.** The cell number density with growth time for three microalgae species.

The PH of the cultures were in 7.5~9.7, 7.5~9.8, 7.5~9.9 for microalgae species *Chlorella vulgaris*, *Chlorella pyrenoidosa*, *Chlorella protothecoides,* respectively. The increase of PH was caused by the growth of the microalgae. Here, the increase of PH was larger than the value reported by Heng and Plion [25]. It was attributed to continuous consumption of $CO_2$ in the culture medium during microalgae growth. Indeed, dissociation of $CO_2$ in the culture medium will cause it becomes slightly acidic.

Figure 5 shows the cell size distribution data of the microalgae species measured at different growth time and *Chlorella protothecoides* is taken as example. The cells are approximately ellipsoid, statistical values of both the major and minor diameters are presented. The cell size distribution is estimated from at least 280 cells for microalgae suspensions at different growth phases. The average cell major and minor diameters didn't vary appreciably with the elapse of growth time. The cell size approximately follows the normal distribution. In all cases, the average cell major and minor diameters are in the ranges 3.93~4.12 and 2.95~3.15, respectively.

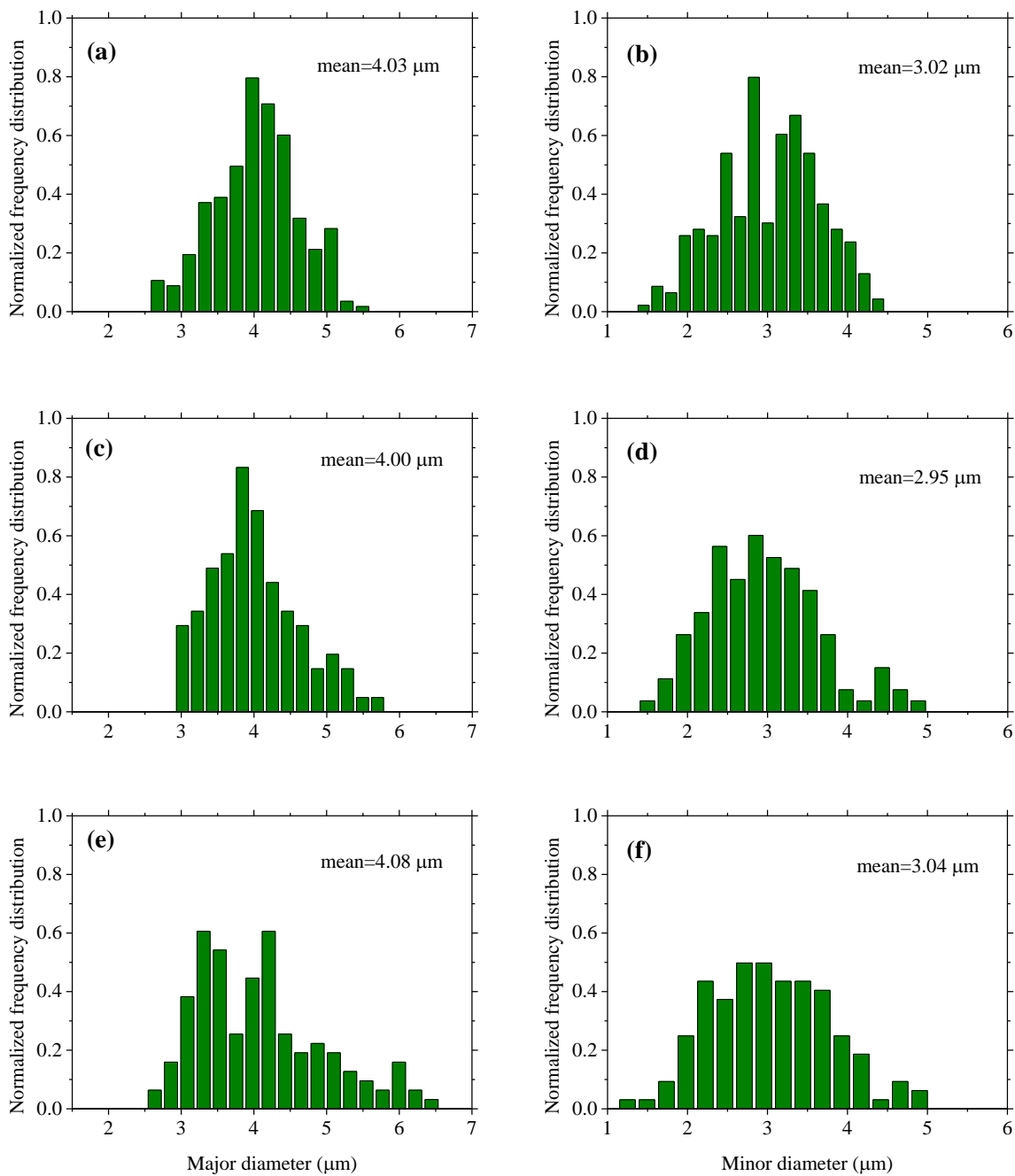

**FIG. 5.** The cell size distribution of the major and minor diameters and the associated mean diameter for *Chlorella protothecoides* at different growth time, (**a**) and (**b**) 4th day, (**c**) and (**d**) 5th day, and (**e**) and (**f**) 8th day.

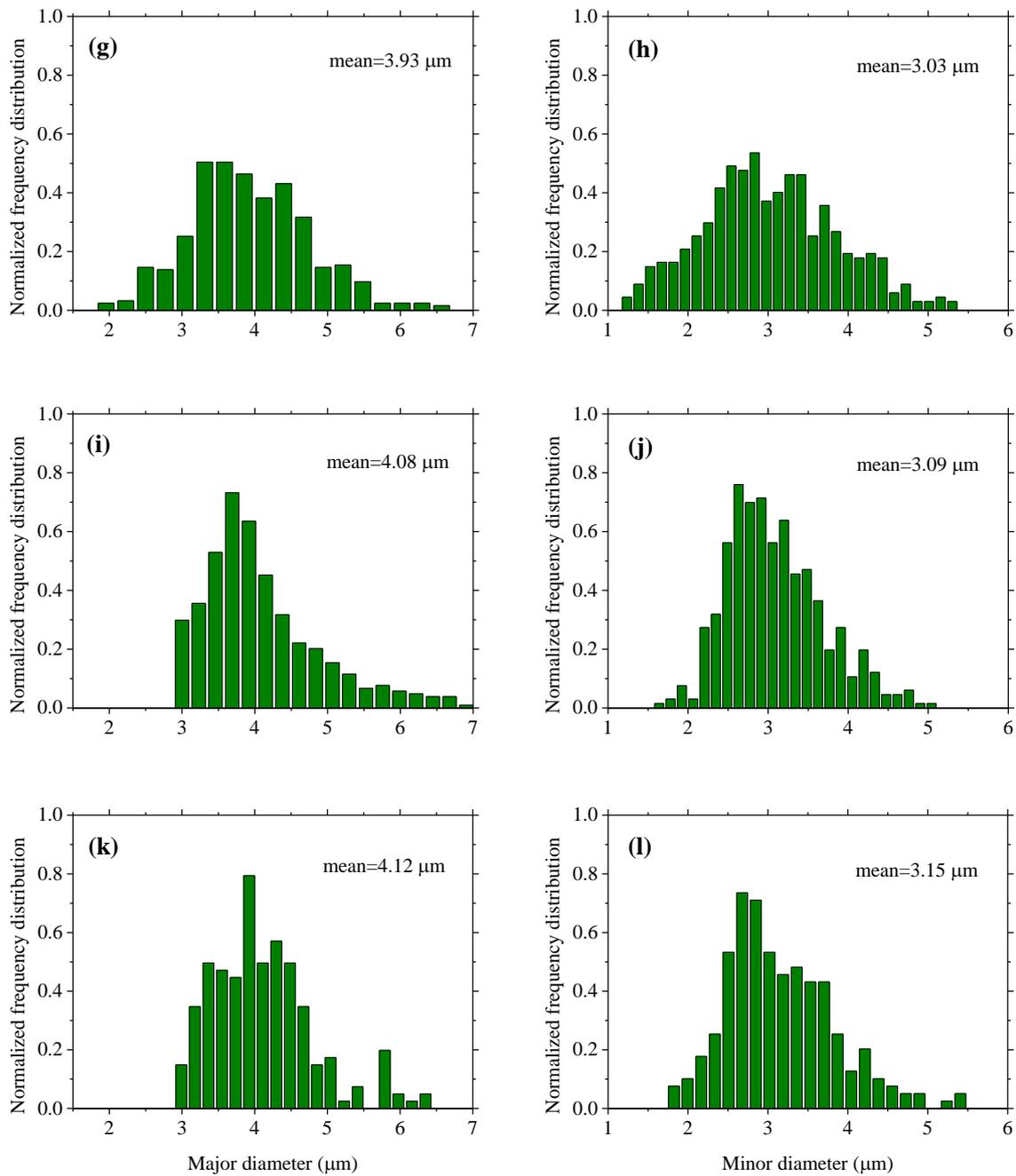

**FIG. 5.** (*continued*) The cell size distribution of the major and minor diameters and the associated mean diameter for *Chlorella protothecoides* at different growth time, (**g**) and (**h**) 9th day, (**i**) and (**j**) 14th day, and (**k**) and (**l**) 18th day.

The measured time dependent spectral absorption and scattering cross-sections of the three microalgae species at the spectral range 380 to 850 nm are presented in Fig. 6. To ease the analysis of temporal dependence behavior, the absorption cross-section of different microalgae species at two specific wavelengths (485 nm and 676 nm) are presented in Fig. 7. As shown in Fig. 6, the scattering cross-section of *Chlorella pyrenoidosa* is generally larger than that of the other two, which is attributed to its relatively larger cell diameter. Three peaks can be observed in the spectra of absorption cross-sections of different microalgae species, i.e., 435 nm, 485 nm, and 676 nm, of which the first and third are corresponding to absorption of chlorophyll a, and the second is corresponding to carotenoids [25]. Some dips in the scattering cross-sections are also observed around the absorption peaks of absorption cross-section, which is due to causality, e.g., the correlation of the real part and imaginary part of the complex refractive index given by the Kramers–Krönig relation [35].

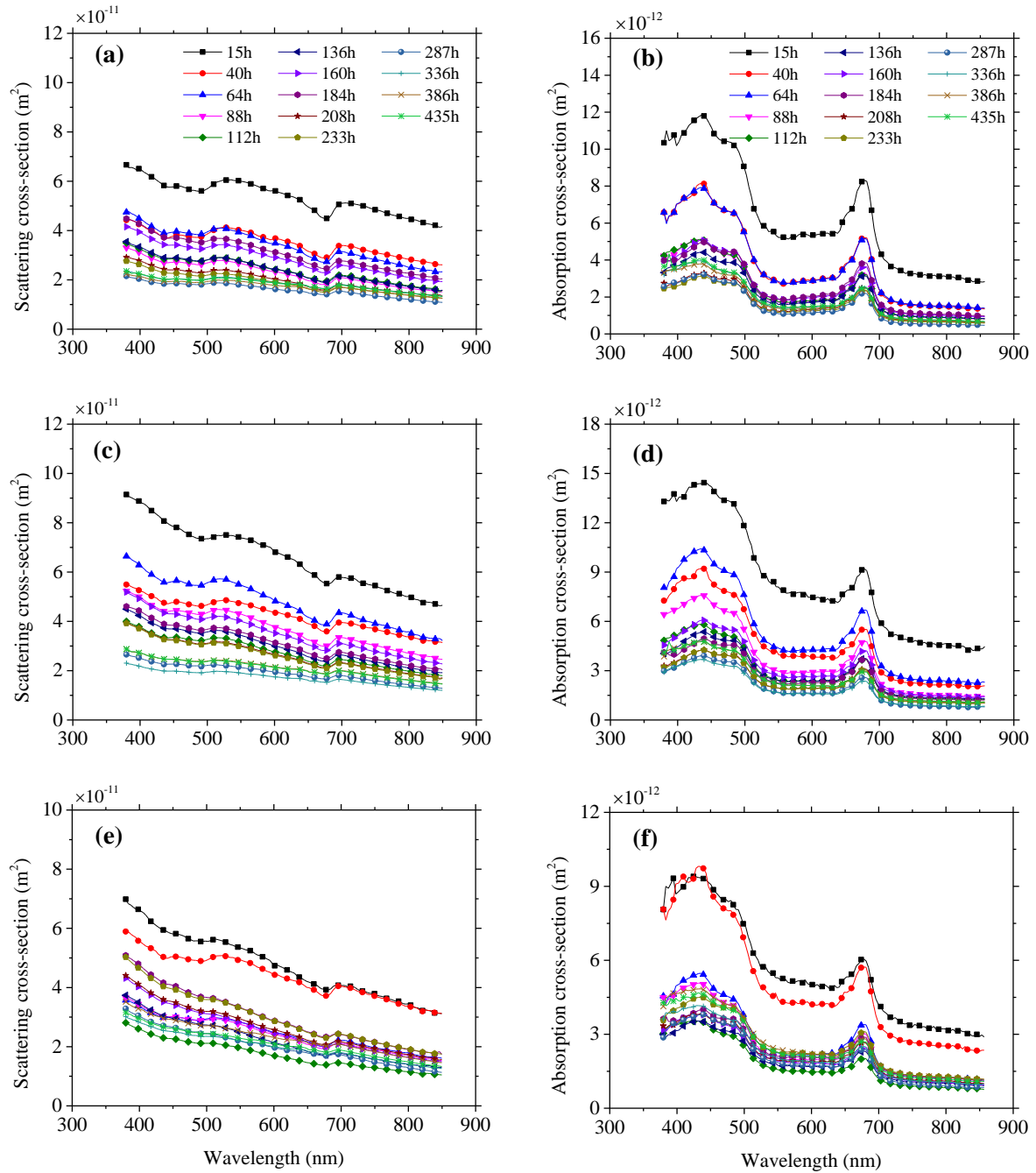

**FIG. 6.** Temporal evolution of spectral scattering and absorption cross-sections of different microalgae species, **(a)** and **(b)** *Chlorella vulgaris*, **(c)** and **(d)** *Chlorella pyrenoidosa*, **(e)** and **(f)** *Chlorella protothecoides*, respectively.

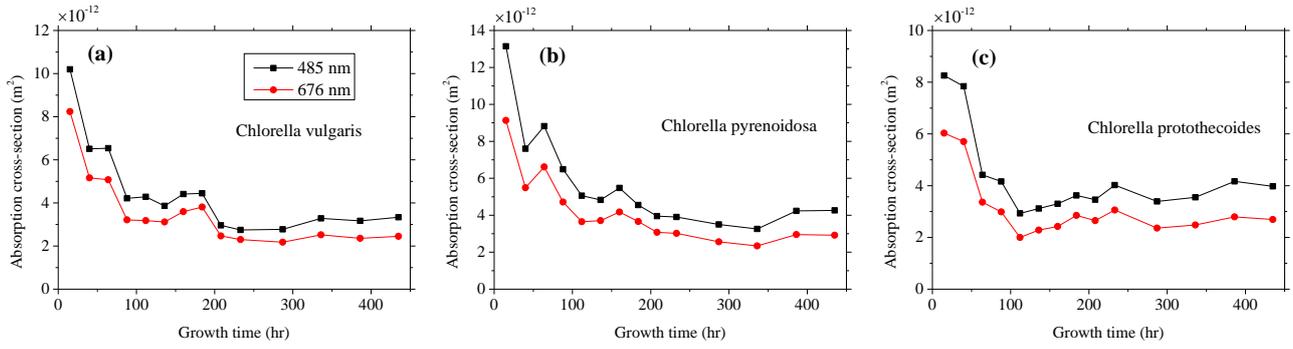

**FIG. 7.** Time dependent absorption cross-sections of different microalgae species at wavelengths 485 nm and 676 nm, **(a)** *Chlorella vulgaris*, **(b)** *Chlorella pyrenoidosa*, and **(c)** *Chlorella protothecoides*, respectively.

The absorption and scattering cross-sections of different microalgae species vary significantly with growth time, e.g., for *Chlorella pyrenoidosa*, the scattering cross-sections decreases from about $9 \times 10^{-11}$ m$^{-2}$ (15h) to about $2 \times 10^{-11}$ m$^{-2}$ (336h) at 380nm, and the absorption cross-sections decreases from about $1.5 \times 10^{-11}$ m$^{-2}$ (15h) to about $3 \times 10^{-12}$ m$^{-2}$ (336h) at 450nm. The larger values are more than 3 times of the smaller value, which indicates large errors will be introduced if only the radiative properties at stationary phase are used for light field analysis in PBRs. The results for the other two microalgae species are similar. As shown in Fig. 7, the absorption cross-sections show largest value at initial time for different microalgae species, then decrease with time and show a local minimum at the exponential growth phase, and then gradually approach to a stable value during the stationary growth phase. Generally, the absorption and scattering cross-sections tend to decrease with the elapse of growth time.

### 5.2 Temporal scaling of the absorption and scattering cross-sections

As shown in Fig. 6, some similarity of the spectra of absorption and scattering cross-sections at different growth period can be observed. A theoretical proof and analysis of this similarity behavior has been presented in Section 3. Here, experimental results of the TSFs of absorption and scattering cross-sections (defined by Eq. (19) and (21)) are presented. Figure 8 shows the TSFs of spectral absorption and scattering cross-sections of the three microalgae species at different growth time. As shown, except the absorption TSFs at initial time (15h), the absorption and scattering TSFs of different microalgae species are nearly independent of wavelength at different growth time. This well confirms the theoretically predicted temporal scaling relation of radiative properties of microalgae established in Section 3, namely, the TFSs of absorption and scattering cross-sections are independent of wavelength, and which are only functions of growth time.

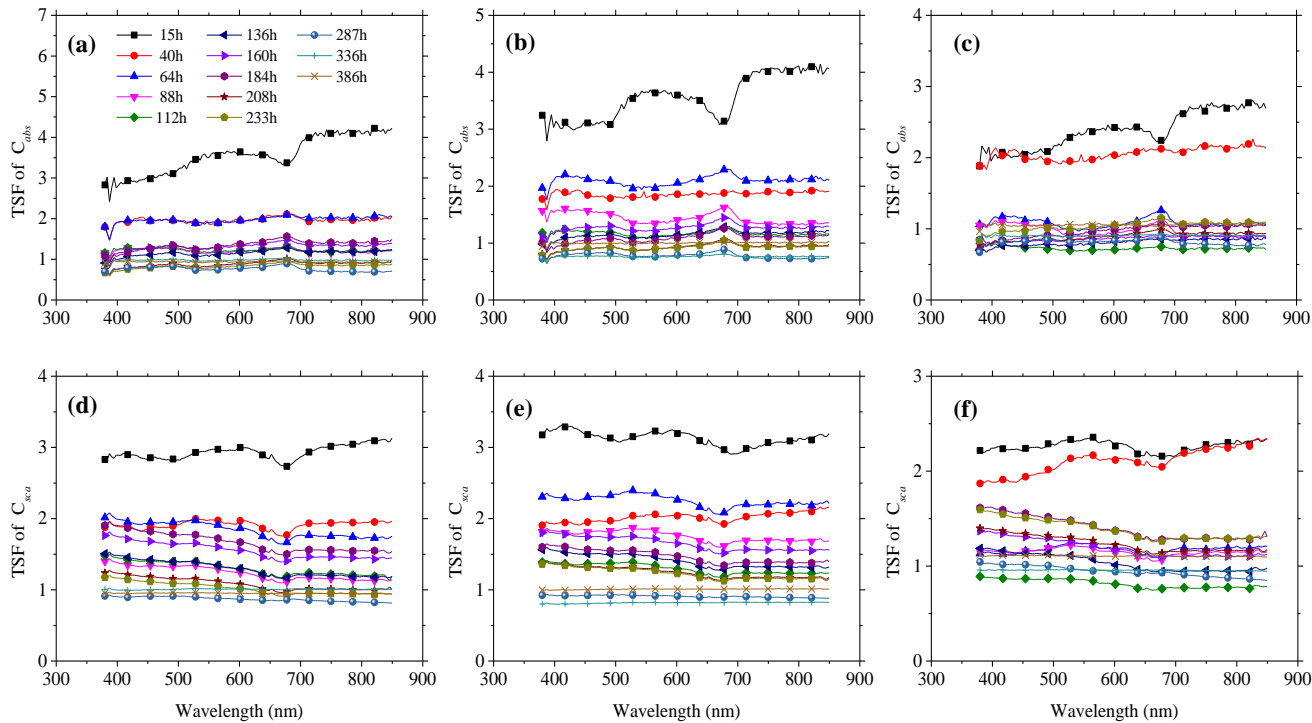

**FIG. 8.** Experimental temporal scaling function (TSF) of absorption and scattering cross-sections for different microalgae species, **(a)** and **(d)** *Chlorella vulgaris*, **(b)** and **(e)** *Chlorella pyrenoidosa*, and **(c)** and **(f)** *Chlorella protothecoides*.

Though obvious spectral features are observed in the absorption cross-sections of different microalgae species, it is interesting to note that the TFSs tend to be independent of wavelength. As such, if the absorption and scattering TSFs of microalgae at different growth period are established, the radiative properties at any growth time can be calculated based on the radiative properties measured at the stationary phase. In the following, the dependence of TSFs on time for different microalgae are presented and analyzed.

Figure 9 shows the spectral averaged TSFs of absorption and scattering cross-sections of the three microalgae species, in which the standard deviations are also given. It is observed that the TSFs of the three microalgae species decrease sharply before 150 hours, which is in the exponential growth phase, and then varies gradually with growth time.

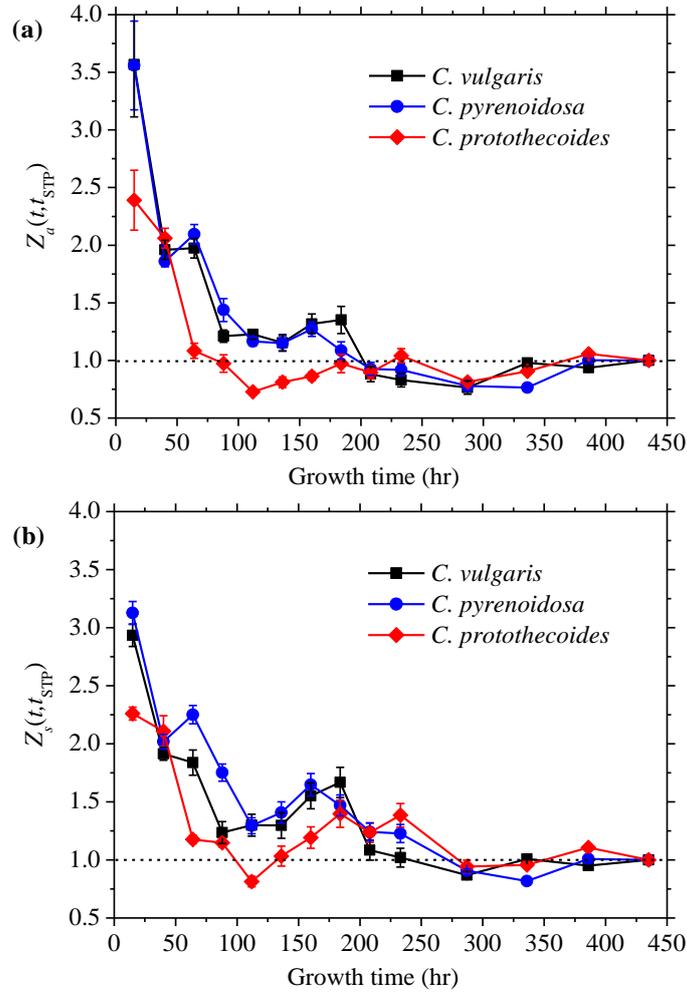

**FIG. 9.** The experimental TSFs of absorption and scattering cross-sections of three microalgae species,

**(a)** TSF of $C_{abs}$, **(b)** TSF of $C_{sca}$.

The decrease of the absorption and scattering TSFs corresponds to the decrease of absorption and scattering absorption cross-sections. From the theoretical analysis (Eqs. (14) and (18)), the decrease of absorption TSF with time indicates the average accumulation rate of pigments in each cell ($\mu_p$) will have a negative value. This is attributed to the cell division that is the dominated biological process in the exponential growth phase, which will result in more smaller microalgae cells, and hence lower average pigment mass concentration in each cell. An analogy with the absorption TSF can be applied to understand the trend of the scattering TSFs. The value of the standard deviations shows the degree of TSF that deviates from a spectral constant value. As shown, the standard deviations of the absorption TSFs of different microalgae species initially have a relatively big value at the start of exponential growth phase (e.g. 15hr), and which tends to decrease with growth time. For the scattering TSFs, there is not a clearly trend for the standard deviations. For all the three microalgae species, the standard deviations are less than 6% of the corresponding TSF of absorption and scattering cross-sections. This is a solid experimental evidence of the temporal scaling relation of the spectral absorption cross-sections and scattering cross-sections of microalgae at different growth period.

## 6. Conclusions

The radiative properties of microalgae will vary with growth time due to several factors, such as the accumulation of pigment and lipid, cell division and metabolism. A temporal scaling relation of the growth dependent radiative properties of microalgae cell suspensions is reported with both experimental observations and theoretical proof. The temporal evolution and temporal scaling characteristics of scattering and absorption cross-sections of three example microalgae were experimentally measured at spectral range 380 to 850 nm. The absorption and scattering cross-sections of the example microalgae species varied significantly with cultivation time, which indicates large errors will be introduced if only the radiative properties at stationary phase are used for light field analysis in PBRs. The TSFs of both the absorption and scattering cross-sections of different microalgae species are approximately constant at different wavelength. Theoretical predictions of the temporal scaling relation are well confirmed by experimental observations. By using the temporal scaling relation expressed with TSFs, the radiative properties at any growth time can be calculated based on the radiative properties measured at the stationary phase. These findings will help the understanding of time dependent radiative properties of microalgae and facilitate their applications in light field analysis in PBRs design. It is noted that this work is still an initial attempt in understanding and modeling of the growth dependent radiative properties of microalgae. The proof of the temporal scaling relation for the scattering cross-sections is by analogy with the absorption cross-sections, though agree with experimental observations, it still lacks theoretical rigidity, hence it needs to be studied furhter.

**Acknowledgements**

This work was supported by National Nature Science Foundation of China (No. 51276051), which is gratefully acknowledged.

**References**

[1] Pilon L, Berberoglu H and Kandilian R. Radiation transfer in photobiological carbon dioxide fixation and fuel production by microalgae. JQSRT 2011; 112: 2639-60.

[2] Parry M, Canziani O, Palutikof J, Van der Linden P and Hanson C. Contribution of working group II to the fourth assessment report of the intergovernmental panel on climate change, 2007. Cambridge, United Kingdom: Cambridge University Press; 2007.

[3] BP. Bp energy outlook 2035. 2015.

[4] Ullah K, Ahmad M, Sharma VK, Lu P, Harvey A, Zafar M and Sultana S. Assessing the potential of algal biomass opportunities for bioenergy industry: A review. Fuel 2015; 143: 414-23.

[5] Hall DO, Markov SA, Watanabe Y and Rao KK. The potential applications of cyanobacterial photosynthesis for clean technologies. Photosynthesis research 1995; 46: 159-67.

[6] Skjånes K, Lindblad P and Muller J. Bioco$_2$–a multidisciplinary, biological approach using solar energy to capture $CO_2$ while producing h$_2$ and high value products. Biomolecular engineering 2007; 24: 405-13.

[7] Yoon JH, Sim SJ, Kim M-S and Park TH. High cell density culture of anabaena variabilis using repeated injections of carbon dioxide for the production of hydrogen. International Journal of Hydrogen Energy 2002; 27: 1265-70.

[8] Ke B. Photosynthesis photobiochemistry and photobiophysics. Springer Science & Business Media; 2001.

[9] Amaro HM, Macedo ÂC and Malcata FX. Microalgae: An alternative as sustainable source of biofuels. Energy 2012; 44: 158-66.

[10] Halim R, Danquah MK and Webley PA. Extraction of oil from microalgae for biodiesel production: A review. Biotechnology advances 2012; 30: 709-32.

[11] Chelf P, Brown L and Wyman C. Aquatic biomass resources and carbon dioxide trapping. Biomass and Bioenergy 1993; 4: 175-83.

[12] Davis R, Aden A and Pienkos PT. Techno-economic analysis of autotrophic microalgae for fuel production. Applied Energy 2011; 88: 3524-31.

[13] Jorquera O, Kiperstok A, Sales EA, Embiruçu M and Ghirardi ML. Comparative energy life-cycle

analyses of microalgal biomass production in open ponds and photobioreactors. Bioresource technology 2010; 101: 1406-13.

[14]Pate R, Klise G and Wu B. Resource demand implications for us algae biofuels production scale-up. Applied Energy 2011; 88: 3377-88.

[15]Lee J-S and Lee J-P. Review of advances in biological $CO_2$ mitigation technology. Biotechnology and Bioprocess Engineering 2003; 8: 354-9.

[16]Ogbonna JC, Soejima T and Tanaka H. An integrated solar and artificial light system for internal illumination of photobioreactors. Journal of Biotechnology 1999; 70: 289-97.

[17]Melis A, Neidhardt J and Benemann JR. Dunaliella salina (chlorophyta) with small chlorophyll antenna sizes exhibit higher photosynthetic productivities and photon use efficiencies than normally pigmented cells. Journal of Applied Phycology 1998; 10: 515-25.

[18]Cornet J, Dussap C and Dubertret G. A structured model for simulation of cultures of the cyanobacterium spirulina platensis in photobioreactors: I. Coupling between light transfer and growth kinetics. Biotechnology and Bioengineering 1992; 40: 817-25.

[19]Cornet J-F, Dussap C, Gros J-B, Binois C and Lasseur C. A simplified monodimensional approach for modeling coupling between radiant light transfer and growth kinetics in photobioreactors. Chemical Engineering Science 1995; 50: 1489-500.

[20]Berberoglu H, Gomez PS and Pilon L. Radiation characteristics of botryococcus braunii, chlorococcum littorale, and chlorella sp. Used for $CO_2$ fixation and biofuel production. JQSRT 2009; 110: 1879-93.

[21]Berberoglu H and Pilon L. Experimental measurements of the radiation characteristics of anabaena variabilis atcc 29413-u and rhodobacter sphaeroides atcc 49419. International Journal of Hydrogen Energy 2007; 32: 4772-85.

[22]Berberoglu H, Pilon L and Melis A. Radiation characteristics of chlamydomonas reinhardtii cc125 and its truncated chlorophyll antenna transformants tla1, tlax and tla1-cw+. International Journal of Hydrogen Energy 2008; 33: 6467-83.

[23]Heng R-L, Lee E and Pilon L. Radiation characteristics and optical properties of filamentous cyanobacterium anabaena cylindrica. Journal of the Optical Society of America A 2014; 31: 836-45.

[24]Li X, Zhao J, Liu L and Zhang L. Optical extinction characteristics of three biofuel producing microalgae determined by an improved transmission method. Particuology 2017; 33: 1-10.

[25]Heng R-L and Pilon L. Time-dependent radiation characteristics of nannochloropsis oculata during batch culture. JQSRT 2014; 144: 154-63.

[26]Modest M. Radiative heat transfer. Academic Press; 2003.

[27]Kandilian R, Lee E and Pilon L. Radiation and optical properties of nannochloropsis oculata grown under different irradiances and spectra. Bioresource technology 2013; 137: 63-73.

[28]Kurano N and Miyachi S. Selection of microalgal growth model for describing specific growth rate-light response using extended information criterion. Journal of Bioscience & Bioengineering 2005; 100: 403-8.

[29]Heng R-L, Lee E and Pilon L. Radiation characteristics and optical properties of filamentous cyanobacterium anabaena cylindrica. JOSA A 2014; 31: 836-45.

[30]Bidigare RR, Smith R, Baker K and Marra J. Oceanic primary production estimates from measurements of spectral irradiance and pigment concentrations. Global Biogeochemical Cycles 1987; 1: 171-86.

[31]Bidigare RR, Ondrusek ME, Morrow JH and Kiefer DA. In-vivo absorption properties of algal pigments. Proceedings of SPIE 1302, Ocean optics X 1990: 290-302.

[32]Pottier L, Pruvost J, Deremetz J, Cornet JF, Legrand J and Dussap C. A fully predictive model for one-dimensional light attenuation by chlamydomonas reinhardtii in a torus photobioreactor. Biotechnology and Bioengineering 2005; 91: 569-82.

[33]Li XC, Zhao JM, Wang CC and Liu LH. Improved transmission method for measuring the optical extinction coefficient of micro/nano particle suspensions. Appl Opt 2016; 55: 8171-9.

[34]Ma CY, Zhao JM, Liu LH, Zhang L, Li XC and Jiang BC. GPU-accelerated inverse identification of radiative properties of particle suspensions in liquid by the Monte Carlo method. JQSRT 2016; 172: 146-59.

[35]Bohren CF and Huffman DR. Absorption and scattering of light by small particles. New York: Wiley; 1983.